\documentclass[a4paper,11pt]{article}
\usepackage{jheppub} 
\usepackage{lineno}
\usepackage{multirow}
\usepackage{booktabs}
\usepackage{subcaption}
\usepackage{tabularx,array,booktabs,multirow,bigstrut}
\usepackage{siunitx}
\usepackage{float}
\usepackage{indentfirst}
\usepackage{amssymb}

\usepackage{comment}

\newcommand{\ttb}{t\bar{t}}

\makeatletter

\newcommand{\Rmnum}[1]{\expandafter\@slowromancap\romannumeral #1@}
\makeatother

\title{\boldmath  Quantum Entanglement between gauge boson pairs at a Muon Collider }

\author[a,b]{Ran Ding,}
\emailAdd{ran.ding@cern.ch}

\author[a]{Alim Ruzi}
\emailAdd{alim.ruzi@pku.edu.cn}

\author[a]{Sitian Qian}
\author[a]{Andrew Levin}
\author[a]{Youpeng Wu}

\author[a]{Qiang Li}
\emailAdd{qliphy0@pku.edu.cn}

\affiliation[a]{
School of Physics and State Key Laboratory of Nuclear Physics and Technology, Peking University, Beijing, 100871, China}
\affiliation[b]{
School of Physics, Sun Yat-Sen University, Guangzhou 510275, China}

\abstract{
Quantum entanglement is one of significant physics phenomena that can be examined at a particle collider. A muon collider can provide a stage on which we can study substantial physics phenomenon, starting from the precision measurements of the Standard Model and beyond to the undiscovered area of physics. In this work, we present a through study of quantum entanglement in $\mu^+\mu^-\to ZZ$ events at a future muon collider. By fixing the spin density matrix, observables quantifying entanglement between $Z$ boson pairs can be measured. After systematic Monte-Carlo simulation and background analysis, we measure the value of entanglement variables and perform hypothesis testing against the non-entangled hypothesis, finally observing the entanglement of the $ZZ$ system up to $2$ significance level.
}

\begin{document}
\maketitle
\flushbottom

\section{Introduction}
Quantum mechanics and special relativity construct the main building blocks of the modern theory of elementary particle interactions, known as the Standard Model (SM) of particle physics. Quantum mechanics, with its probabilistic nature, provides our understandings of the physics law in the microscopic world of particles. One of the distinctive nature of this theory is that whether it contradicts the predictions of the local realistic theories. In recent years, we have witnessed breakthroughs in the study of quantum entanglement (QE) and quantum information science at the high energy frontier. For example, several works show that violation of Bell Inequality could be measured in the $\ttb$ system at the Large Hadron Collider (LHC)~\cite{Afik:2020onf,Fabbrichesi:2021npl,Afik:2022dgh, Severi:2021cnj,Aguilar-Saavedra:2022uye,Aoude:2022imd,Afik:2022kwm,Han:2023fci,Dong:2023xiw,Severi:2022qjy,Maltoni:2024csn,Aguilar-Saavedra:2024hwd,Aguilar-Saavedra:2024vpd,Cheng:2024btk}. Considering the heavy mass and very short life time, the $\ttb$ system can be an ideal platform to perform quantum tests at the LHC. Both ATLAS and CMS collaborations~\cite{ATLAS:2023fsd,CMS:2024pts,CMS:2024zkc} have measured entanglement among top quark pairs with high sensitivity. Up to now, search for QE and test Bell nonlocality between other pair of system have continuously gained growing interests at collider experiment.

Quantum state tomography~\cite{White:1999sjn,James:2001klt,Thew:2002fom}, determining the density matrix from an ensemble of measurements, has emerged as a cornerstone technique for reconstructing the quantum state of a system, providing invaluable insights into quantum correlations, coherence, and entanglement~\cite{Popescu:1994kjy, Einstein:1935rr,Bell:1964kc,Barr:2024djo,Martens:2017cvj,Bernal:2023jba}. While initially developed within the realm of quantum optics and low-dimensional systems, advancements in experimental and theoretical physics have extended its applicability to high-energy particle physics, particularly in exploring the quantum properties of massive particles produced at particle colliders~\cite{Aguilar-Saavedra:2015yza,Aguilar-Saavedra:2017zkn,Aguilar-Saavedra:2022mpg,Aguilar-Saavedra:2022wam,Ashby-Pickering:2022umy,Barr:2021zcp,Larkoski:2022lmv,Fabbrichesi:2024wcd,Ma:2023yvd,Fedida:2022izl,Fabbrichesi:2023jep,Han:2025ewp,Cheng:2025cuv,Cheng:2024rxi,Han:2024ugl}. The advent of high-energy colliders such as LHC and the proposed muon collider opens a promising avenue for probing fundamental aspects of quantum mechanics in previously inaccessible regimes.

The study of massive spin-1 particles, such as the electroweak gauge bosons $WW$ and $ZZ$, is particularly compelling in this context. These particles play a critical role in the Standard Model of particle physics, mediating weak interactions and participating in processes that probe electroweak symmetry breaking~\cite{Glashow:1961tr,Higgs:1964pj,Salam:1964ry,Weinberg:1967tq}. At a muon collider~\cite{AlAli:2021let, Boscolo:2018ytm}, where precise beam properties and high luminosity enable clean experimental conditions, the production of such particles offers an unparalleled opportunity to investigate their quantum properties. Specifically, quantum state tomography of $WW$ and $ZZ$ boson pairs provides a direct method to study their spin correlations, polarization states, and entanglement, enriching our understanding of the quantum nature of the Standard Model.

In this article we use quantum state tomography of two massive spin-1 particles at a muon collider to fix the spin density matrix from a simulation events of $\mu^+\mu^-\to ZZ\to4\ell$. Quantum observables like concurrence are also measured to examine whether the $ZZ$ pairs are entangled or not at extremely relativistic environment. By leveraging the unique features of the muon collider—such as its clean experimental environment, minimal hadronic background, and capability to operate at high center-of-mass energies, we aim to explore the quantum properties inside $ZZ$ boson pairs with high signal purity.

\section{Quantum state tomography for massive gauge boson pairs}

\subsection{spin density  matrix}
\label{sec:tomography}
As the most important object in this study, the density matrix~\cite{Fano:1957zz} is crucial for understanding and analyzing the characteristics of a quantum system, defined as
\begin{equation}
    \label{eq:def_rho_init}
    \rho=\sum_i p_i\left|\psi_i\right\rangle\left\langle\psi_i\right|,
\end{equation}
where $\left\{\psi_i\right\}$ is a set of pure states, $p_i$ is the classical probability that satisfies $p_i\geq 0$ with $\sum_i p_i=1$.

In practical applications, we focus primarily on the spin density  matrix of a system. For a single particle with spin $j$, its spin density  matrix is an operator on its (2$j$ + 1)-dimensional spin Hilbert space. In the case of multi-particle systems, the density matrix acts on the tensor product space $\mathcal{H} = \mathcal{H}_1 \otimes \mathcal{H}_2 \otimes \ldots \otimes \mathcal{H}_n$ of the spin states of each subsystem.

The spin density  matrix for a pair of spin-1 massive particles generated through an interaction can be derived by calculating its polarized scattering amplitude in the helicity basis as 
\begin{equation}
     \rho _{\lambda_1 \lambda_1', \lambda_2, \lambda_2'} = \frac{\mathcal{M}_{\lambda_1\lambda_2} \mathcal{M}_{\lambda_1'\lambda_2'}^{\dagger}}{\left|\mathcal{M}\right|^2} ,
\end{equation}
where $|\mathcal{M}|^2$ is the unpolarized squared amplitude, $\lambda_1$ and $\lambda_2$ are the helicity basis states of the massive particle pair. This corresponds to defining the spin quantization axis along the momentum direction. 

Since the polarization information of the massive spin-1 particle cannot be directly obtained in collider experiments, one approach is to extract its spin density matrix through quantum state tomography. This is usually done by measuring the angular distribution of the decay products, such as leptons.
The spin density matrix is typically parameterized using the generalized d-dimensional Gell-Mann operator. For a single massive spin-1 particle, the density matrix can be decomposed into the eight Gell-Mann matrices $T_i$ and the $3$-dimensional identity matrix $I_3$, as~\cite{Fabbrichesi:2023cev}
\begin{equation}
    \rho = \frac{1}{3}I_3 + \sum^{8}_{i=1} a_i T_i,
\end{equation}
where $a_i$ are real coefficients. 

Since the Gell-Mann matrices satisfy $\text{Tr}\left[T_iT_j\right]=2\delta_{ij}$ and $\text{Tr}\left[T_i\right]=0$, once the form of the density matrix is determined, $a_i$ can be obtained by the following equation:
\begin{equation}
    a_i = \frac{1}{2}\text{Tr}\left[\rho T_i\right].
\end{equation}
Taking a system of two massive spin-1 particles as an example, the spin density matrix can be expressed as
\begin{equation}
    \label{eq:generaldec}
    \rho = \frac{1}{9} I_9 + \sum_{i=1}^{8} A_i T^i \otimes I_3 
+ \sum_{j=1}^{8} B_j I_3 \otimes T^j 
+ \sum_{i=1}^{8} \sum_{j=1}^{8} C_{ij} T^i \otimes T^j,
\end{equation}
where $A_i$ and $B_i$ parametrize the individual polarizations of each particle, while $C_{ij}$ encodes the their "correlation". 
Similarly, these coefficients can be obtained by the following relations:
\begin{equation}
    \begin{aligned}
        A_i = \frac{1}{6}\text{Tr}\left[\rho T_i \otimes I_3 \right],\quad
        B_i = \frac{1}{6}\text{Tr}\left[\rho I_3 \otimes T_i\right],\quad
        C_{ij} = \frac{1}{4}\text{Tr}\left[\rho T_i \otimes T_j\right].
    \end{aligned}
\end{equation}

\subsection{Extraction of density matrix from data}
\label{sec:tomography_method}
The extraction of $\rho$ refers to determining the coefficients in Eq.~\ref{eq:generaldec} from experimental data.
As rigorously established in Ref.~\cite{Ashby-Pickering:2022umy}, the general procedure is implemented through 
the Wigner-Weyl formalism between functions in the classical phase space and the operators in the Hilbert space.
The basic idea is to measure the momentum $\mathbf{p}$ of the decay products in the parent particle's rest frame, which is assumed to be the major directly measurable quantity in this study.

We begin by measuring the direction $\hat{\textbf{n}}=\hat{\textbf{n}}\left(\theta,\phi\right)$ of the decay leptons ($\ell^\pm$ for $W^{\pm}$, and $\ell^+$ for $Z$) in the rest frame of their parent particles. The polar axis is defined as the momentum direction of one of the parent particles in the collider center-of-mass frame. Specifically, for a single $W^{\pm}$ decaying into two massless leptons, the density matrix coefficients $a_i$ can be extracted through averaging the eight functions $p_i$, as
\begin{eqnarray}
    \label{eq:calc_a_i}
    \hat{a}_i = \frac{1}{2} \left\langle p_i^\pm(\hat{\textbf{n}}) \right\rangle,
\end{eqnarray}
where $p_i^\pm$ is the Wigner P symbols for the $W^\pm$ boson corresponding to the Gell-Mann operators. Their explicit form is provided in \ref{eq:def_p_plus} of the Appendix.

Unlike $W^\pm$, the coupling between $Z$ and leptons contains both left- and right-handed components. This requires modifying the formalism through the definition:
\begin{eqnarray}
    \tilde{p}_i = \sum_{j=1}^{8}A^j_ip^+_j ,
    \label{eq:calc_p_tilde}
\end{eqnarray}
where the matrix $A_i^j$ is given in \ref{eq:def_A_matrix}, depending on the left- and right-hand coupling $g_R=-\sin ^2 \theta_\text{W}$ and $g_L=1/2 - \sin ^2 \theta_\text{W}$.
Then the parameters of the density matrix are obtained as the expectation value of  the generalized $\tilde{p}$ functions, as 
\begin{eqnarray}
    \hat{a}_i = \frac{1}{2} \left\langle \tilde{p}_i(\hat{\mathbf{n}}) \right\rangle,
\end{eqnarray}
The coefficients of the di-boson system can be calculated by
\begin{eqnarray}
    \begin{aligned}
        \hat{A}_i &= \frac{1}{3} \hat{a}_i = \frac{1}{6} \left\langle p_i^1(\hat{\mathbf{n}}_1) \right\rangle, \quad 
        \hat{B}_i = \frac{1}{3} \hat{b}_i = \frac{1}{6} \left\langle p_i^2(\hat{\mathbf{n}}_2) \right\rangle, \quad
        \hat{C}_{ij} = \frac{1}{4} \left\langle p_i^1(\hat{\mathbf{n}}_1) p_j^2(\hat{\mathbf{n}}_2) \right \rangle.
    \end{aligned}
\end{eqnarray}
And specifically for the $ZZ$ system, since the two $Z$ bosons are indistinguishable, we need to impose an additional symmetry constraint by exchanging the labels $i$ and $j$, as 
\begin{eqnarray}
    \begin{aligned}
        \hat{A}_i &=\hat{B}_i= \frac{1}{12} \left\langle p_i^1(\hat{\mathbf{n}}_1)+p_i^2(\hat{\mathbf{n}}_2)  \right\rangle, \quad \\
        \hat{C}_{ij}& = \frac{1}{8} \left\langle p_i^1(\hat{\mathbf{n}}_1) p_j^2(\hat{\mathbf{n}}_2)  + p_j^1(\hat{\mathbf{n}}_1) p_i^2(\hat{\mathbf{n}}_2)\right \rangle .
    \end{aligned}
\end{eqnarray}

\section{Observables for quantum entanglement}
\label{sec:observables}
Entanglement and separability are fundamental attributes of quantum systems. In our current research, we primarily focus on the entanglement between the third components of spin for the particles in the final-state system generated in $\mu\mu\to ZZ$ scattering.

Based on the method described in Sec.~\ref{sec:tomography_method}, we can reconstruct the density matrix of the particle systems at a high-energy muon collider through angular distribution measurements of their decay products. The reconstructed density matrix, or equivalently, its Gell-Mann coefficients, enables rigorous quantification of the QE in the system.

For any general bipartite mixed state, if its spin density  matrix can be written as the convex combination of the density matrices of its subsystems, as
\begin{eqnarray}
    \label{eq:rho_sep}
    \rho_{\rm sep} = \sum_i p_i ~\rho_i^A \otimes \rho_i^B,
\end{eqnarray}
the entire state is considered to be separable; otherwise, it is entangled, where $p_i$ is the classical probabilities~\cite{PhysRevA.40.4277}. However, determining separability directly from this definition is computationally intractable, as the problem is provably NP-hard~\cite{gurvits2003classical}.

One of the most popular tests to determine a density matrix is the Peres-Horodecki criterion, also called the Positive Partial Transpose (PPT) criterion, which is a necessary and sufficient condition for determining the presence of entanglement in $2\otimes 2$ or $2\otimes 3$ systems~\cite{PhysRevLett.77.1413,HORODECKI19961}. Given a density matrix, the PPT criterion states that only if its partial transpose matrix, i.e.,
\begin{eqnarray}
    \rho^{\mathrm{T}_2} _{\lambda_1 \lambda_1', \lambda_2, \lambda_2'} = \rho _{\lambda_1 \lambda_1', \lambda_2', \lambda_2},
\end{eqnarray}
is not positive semi-definite (i.e., has at least one negative eigenvalue), the density matrix is entangled; otherwise, it is separable.
To quantify the degree of entanglement, the Negativity is defined by~\cite{PhysRevA.65.032314}
\begin{eqnarray}
    \mathcal{N}(\rho)=\frac{1}{2}\sum_k \left(\left|\mu_k^{\mathrm{T}_2}\right|-\mu_k^{\mathrm{T}_2}\right),
\end{eqnarray}
where $\mu ^{T_2}_k$ is the eigenvalue of $\rho^{\mathrm{T}_2}$. When a significantly non-zero value of $\mathcal{N}$ is observed, it can be concluded that the observed system is entangled at a high confidence level. Note that the PPT criterion is both necessary and sufficient for $2\otimes 2$ or $2\otimes 3$ systems, but only a sufficient condition for entanglement in the general $3\otimes 3$ or higher-dimensional systems.

Another commonly used relevant observable for quantifying QE is known as the concurrence, which can be defined by~\cite{PhysRevLett.98.140505,Rungta:2001zcj}
\begin{equation}
    \mathcal{C}(\rho)=\sqrt{2\left(1-\operatorname{Tr}\left[\left(\rho_A\right)^2\right]\right)}
\end{equation}
for the pure state, where $\rho_A=\text{Tr}_B[\rho]$ is the reduced density matrix of $\rho$. For mixed states, it can be extended as
\begin{eqnarray}
    \mathcal{C}(\rho)=\inf _{\{|\psi_i\rangle\}} \sum_i p_i \mathcal{C}\left(\left|\psi_i\right\rangle\right),
\end{eqnarray}
where the infimum is taken over all the possible pure state decompositions of $\rho$. A non-vanishing concurrence ($\mathcal{C}>0$) serves as a quantitative indicator for bipartite entanglement of the system. However, due to the difficulty of extracting the infimum over all pure state decompositions, the concurrence for the diqutrit systems cannot be analytically calculated. Compared to examining $\mathcal{C}$, the lower bound of $\mathcal{C}^2$ is easier to obtain, as~\cite{PhysRevLett.98.140505}
\begin{eqnarray}
    \left[\mathcal{C}(\rho)\right]^2 \geq c^2_{\text{MB}} = 2~\text{Tr}\left[\rho^2\right] - \text{Tr}\left[(\rho_A)^2\right] - \text{Tr}\left[(\rho_B)^2\right].
\end{eqnarray}
 For a diqutrit system, it reads:
\begin{eqnarray}
    \label{eq:c2_calc}
    c^2_{\text{MB}}=-\frac{4}{9}-6\sum_i A_i^2 - 6\sum_i B_i^2 + 8\sum_{ij}C_{ij}^2,
\end{eqnarray}
where $A_i$, $B_i$ and $C_{ij}$ are the coefficients of the density matrix under the Gell-Mann parameterization. If $c^2_{\text{MB}}>0$, it implies that $\mathcal{C}>0$ and the density matrix $\rho$ represents a entangled state.
    
\section{Application to $\mu^+\mu^-\to$ ZZ states}

\subsection{Signal and background selection}
The $Z$ boson serves as a ideal experimental probe for characterizing qutrit entanglement in collider experiments, owing to its clean experimental signature in charged lepton decay channels, which allows full kinematic reconstruction of multi-$Z$ systems, thereby enabling quantum tomography.

\begin{figure}[!h]
    \centering
    \subfloat[]
    {\includegraphics[width=.5\textwidth]{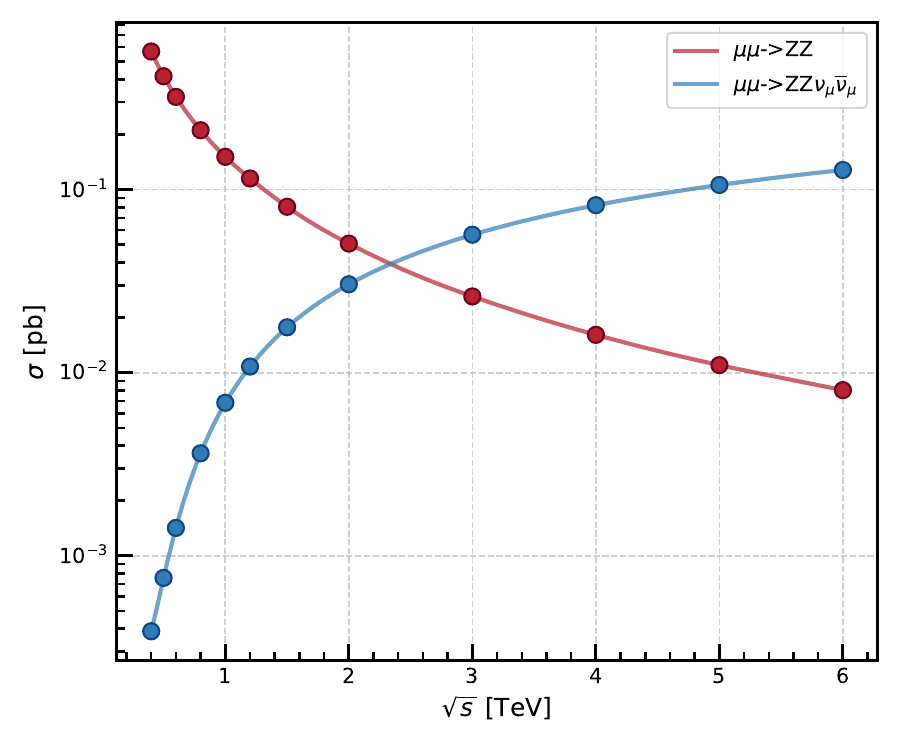}\label{fig:id1_1}}
    \subfloat[]
    {\includegraphics[width=.5\textwidth]{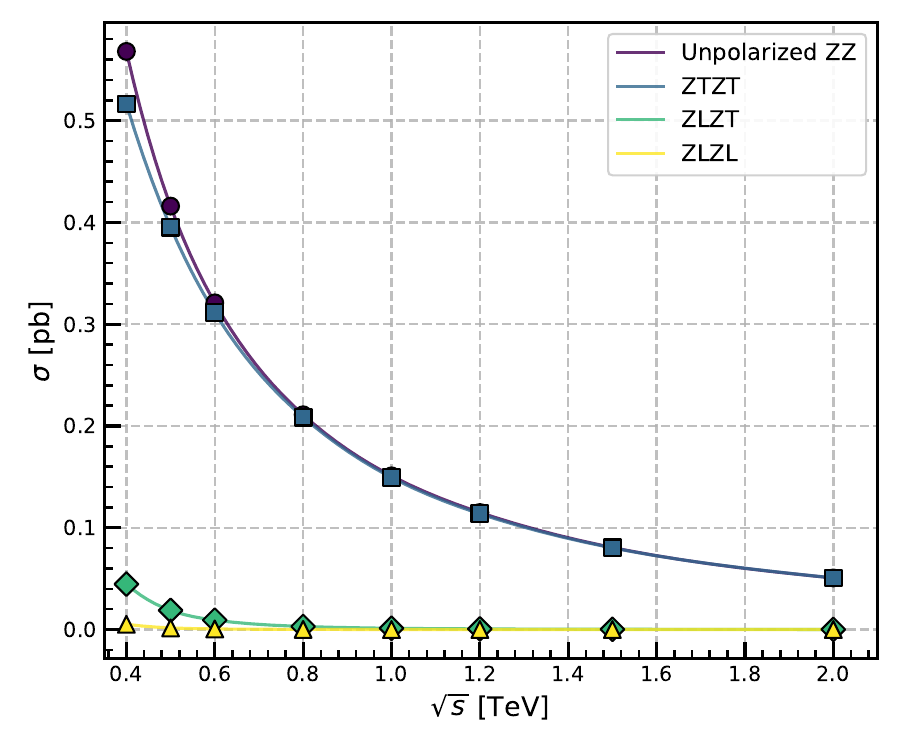}\label{fig:id1_2}}
    \caption{cross-section of $\mu ^+ \mu ^- \to Z Z$ and $\mu ^+ \mu ^- \to Z Z \nu_\mu \bar{\nu}_\mu$ (left) and different polarized combinations of $\mu ^+ \mu ^- \to Z Z$ (right) as a function of the center-of-mass energy.}
    \label{fig:id1}
\end{figure}

Considering the processes of producing $Z$ boson pairs at the muon collider, the $ZZ$ system with the highest degree of entanglement is expected to arise from the Higgs decay, where the Higgs boson mainly comes from $\mu ^+ \mu ^-$ scattering with a pair of neutrinos. This process has been studied in detail in Ref.~\cite{Ruzi:2024cbt}, where the results show that with the luminosity of 30 ab$^{-1}$, the significance for observing QE can reach 4$\sigma$, and the violation of the Bell inequality is about 2$\sigma$. Additionally, other typical $ZZ$ production processes mainly include the direct scattering production process $\mu ^+ \mu ^- \to Z Z$ and the vector boson scattering (VBS) process $\mu ^+ \mu ^- \to Z Z \nu_\mu \bar{\nu}_\mu$. The former is more prevalent at lower center-of-mass energies, while the latter becomes dominant at higher center-of-mass energies, as shown in Fig.~\ref{fig:id1_1}.

\begin{figure}[!h]
\centering
\subfloat[]
{\includegraphics[width=.3\textwidth]{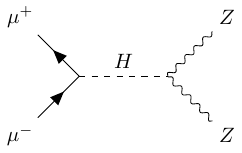}\label{fig:id2_1}}
\subfloat[]
{\includegraphics[width=.3\textwidth]{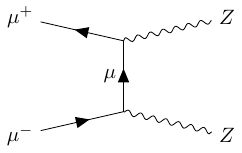}\label{fig:id2_2}}
\subfloat[]
{\includegraphics[width=.3\textwidth]{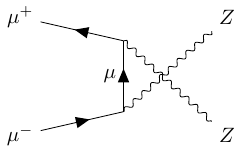}\label{fig:id2_3}}
\caption{The leading order Feynman diagrams of $\mu ^+ \mu ^- \to Z Z$: (a) induced by an off-shell Higgs boson; (b) and (c) induced by the muon in the t-channel and u-channel, respectively.}
\label{fig:id2}
\end{figure}

In this work, we focus on the process $\mu ^+ \mu ^- \to Z_1 Z_2,~ Z_1 \to \ell^+_1\ell^-_1,~Z_2 \to \ell^+_2\ell^-_2$ where $\ell=e,\mu$, considering an experiment conducted at the muon collider with $\sqrt{s}=1~$TeV, and a luminosity $L$ ranging from 5 to 75 ab$^{-1}$. The cross-section contributions from different $ZZ$ polarized states (without including the branching ratio for $Z\to \ell^+\ell^-$) are shown in Fig.~\ref{fig:id1_2}. The leading-order Feynman diagrams are shown in Fig.~\ref{fig:id2}. At a 1 TeV muon collider, the cross-section of Fig.~\ref{fig:id2_1} is extremely low and can essentially be considered as entirely arising from Fig.~\ref{fig:id2_2} and Fig.~\ref{fig:id2_3}.

In the experiment, we search for processes containing two pairs of opposite-charged leptons in the final state, with the requirement that the invariant mass of the lepton pairs to be close to the Z boson mass. Consequently, potential standard model background includes any processes that featuring two pairs of leptons, along with any number of neutrino pairs, jets, and photons in the final state. In the preliminary background study, we do not consider other potential backgrounds beyond the standard model background, such as those arising from particle misidentification or other accidental coincidences, since most of thees have extremely low contributions and can be easily suppressed by applying cuts on the missing energy.

Following Ref.~\cite{Costantini:2020stv}, we categorize the background processes as shown below, in accordance with several previous studies on the muon collider $ZZ$ processes~\cite{Yang:2021zak,Jiang:2024wwa}:
\begin{itemize}
    \item s-channel processes: $\mu^+\mu^- \to X$,
    \item $WW$ fusion: $\mu^+\mu^- \to X\nu_{\mu}\bar{\nu}_{\mu}$,
    \item $ZZ/Z\gamma/\gamma\gamma$ fusion: $\mu^+\mu^- \to X\mu^+\mu^-$,
    \item $WZ$ fusion: $\mu^+\mu^- \to X\mu^-\bar{\nu}_{\mu} (\mu^+\nu_\mu)$,
\end{itemize}
where $X = a~t\bar{t}+b~V+c~\text{H}$, with $a,b,c$ as integers, indicating the number of corresponding components in each process. Each category in the list includes multiple processes and their interferences, but in the current simulation, we only consider a subset of the processes involved, as listed in Tab.~\ref{tab:background_process}.

\begin{table}[h]
\centering
\caption{Summary of the background processes in this study}
\label{tab:background_process}
\begin{tabular}{c c c c }
\toprule
\midrule
Category&Including background processes\\
\midrule
s-channel&$Zt\bar{t}$, $ZWW$, H$t\bar{t}$, $Z$H\\
$WW$ fusion&H, $Z$H, $ZZ$, $ZWW$\\
$ZZ/Z\gamma/\gamma\gamma$ fusion&$WW$, H, $t\bar{t}$\\
$WZ$ fusion&H$W$, $WZ$, $Wt\bar{t}$\\
\midrule
\bottomrule
\end{tabular}
\end{table}

\subsection{Event simulation and estimation}
\label{sec:sbstudy}
In this study, the production cross-sections and corresponding Monte Carlo (MC) event samples for both signal and background processes are generated using MadGraph5\_aMC@NLO (MG) v3.1.1~\cite{Madgraph}. The MG simulation incorporates generator-level cuts, requiring $p_T > 10~\mathrm{GeV}$, $|\eta| < 2.5$, and $\Delta R > 0.4$ for all final-state leptons. Subsequent parton showering and hadronization are carried out using \textsc{Pythia8} (PY8)~\cite{Pythia} with its default models. The detector response is then simulated using \textsc{Delphes}~\cite{Delphes} v3.5.1, employing the built-in muon collider detector card. 

After detector simulation, the collision's center-of-mass frame and both Z boson decay rest frames are reconstructed via four-lepton kinematics, extracting angular distributions for quantum tomography in Sec.~\ref{sec:tomography}. The signal selection requires exactly four charged leptons in the final state with zero lepton flavor. In the $2e2\mu$ channel, the invariant masses of the $e^+e^-$ and $\mu^+\mu^-$ pairs directly correspond to the masses of the two Z bosons. In the 4$e$ or 4$\mu$ channel, we evaluate two opposite-sign lepton pair combinations and compute the total mass deviation, defined as $\Delta M_{\text{tot}}$, as $\Delta M_{\text{tot}} = \sum_{i=1,2}\left|M_{Z}-M_{Z_i}\right|$, where $M_{Z}$ is the mass of the $Z$ boson (91.2~GeV), and $M_{Z_i}$ is the reconstructed dilepton mass. The pairing that minimizes $\Delta M$ is retained for further analysis.

\begin{figure}[h]
    \centering
    \subfloat[$M_{Z_i}$]
    {\includegraphics[width=.5\textwidth]{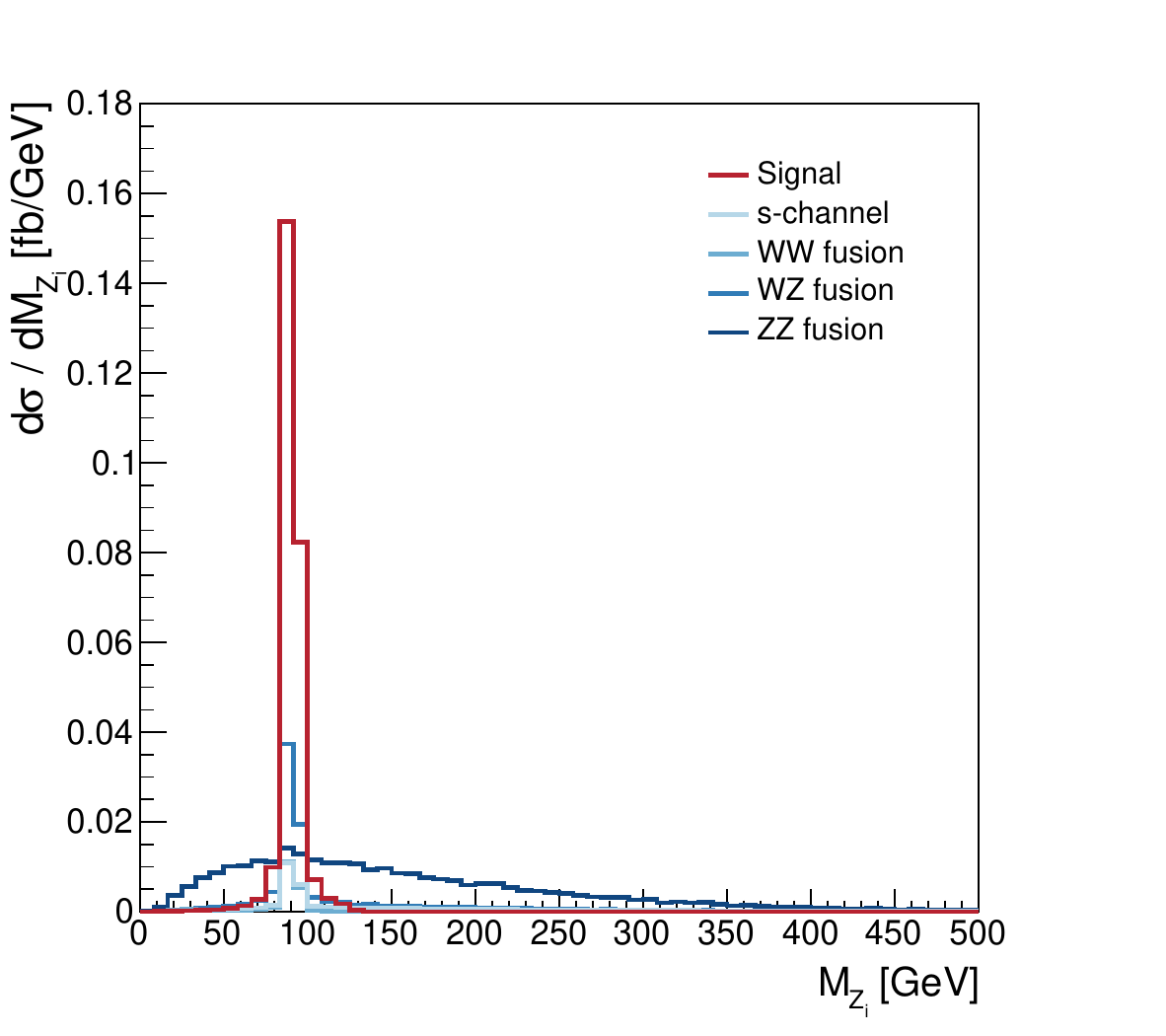}\label{fig:id7_1}}
     \subfloat[$M_{4\ell}$]
    {\includegraphics[width=.5\textwidth]{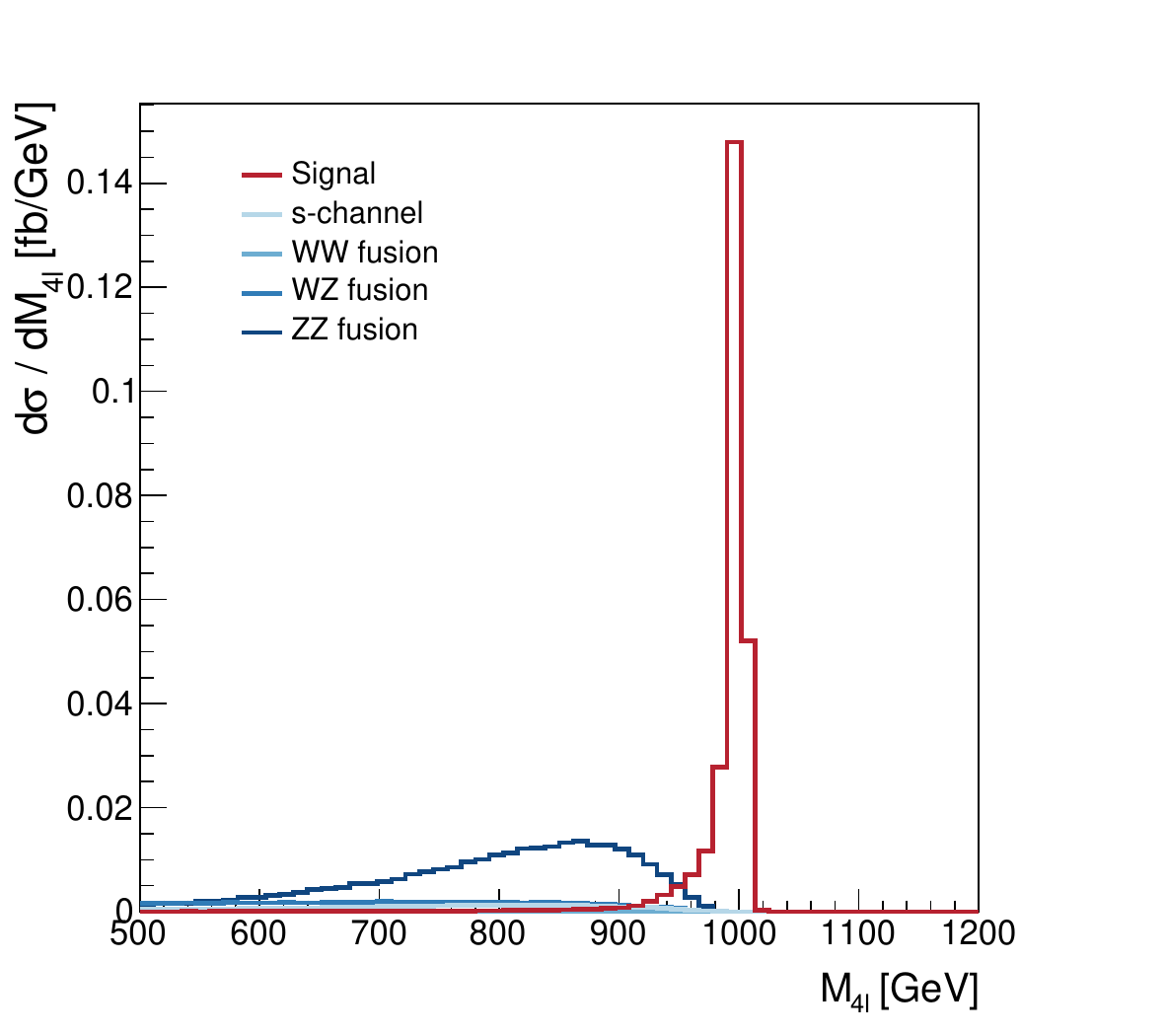}\label{fig:id7_2}}
    \caption{Distributions of $M_{Z_i}$ and $M_{4\ell}$ of signal and specific background processes in Tab.~\ref{tab:background_process}. All histograms are scaled to their corresponding cross-sections.}
    \label{fig:id7}
\end{figure}

Following signal selection, we focus on the pure signal channel $\mu^+\mu^-\to ZZ \to 4\ell$ and compare it to the background processes listed in Tab.~\ref{tab:background_process}. We analyze the reconstructed Z boson mass ($M_{Z_i}$) and the tetra-lepton invariant mass ($M_{4\ell}$) distributions, as shown in Fig.~\ref{fig:id7}. 
The signal process shows distinct mass resonances, with the $M_{Z_i}$ peak at 91.2 GeV, aligned with the nominal Z mass, and the $M_{4\ell}$ distribution centered around the center-of-mass energy (1~TeV).
For the background processes, the $M_{4\ell}$ distributions lies significantly below 1~TeV due to missing energy from undetected particles. Meanwhile, the $M_{Z_i}$ distributions displays a hybrid structure, combining a non-resonant continuum background with a Z-mass peaking component.

Based on these distributions, we apply the following cuts to suppress the background: (a) $86<M_{Z_i}<96$~GeV; (b) $M_{4\ell}>960$~GeV. Then the cut flow and the final selection efficiency are summarized in Tab.~\ref{tab:efficiency_table}, where $\sigma$ is the cross-section, $\varepsilon_{\text{abs}}$ and $\varepsilon_{\text{rel}}$ are the absolute and relative efficiency, respectively. The results shows that after the full signal selection and background suppression, the residual background contribution is negligible, and thus no background is considered in the following analysis.

Notably, a rigorous signal simulation requires consideration of all scattering channels contributing to the $4\ell$ final state, which encompasses both the double $Z$ boson resonant production, and the non-resonant contributions. However, our results reveals that after applying the $M_{Z_i}$ constraint, the non-resonant contributions only exhibits tiny effect on the extracted density matrix. Specifically under realistic experimental statistics, their contribution becomes substantially smaller than the statistical uncertainties. Therefore, the signal simulation in this study is limited to $\mu^+\mu^-\to ZZ\to 4\ell$.

\begin{table}[htbp]
    \centering
    \renewcommand{\arraystretch}{1.3}
    \begin{tabular}{l c c c c c c}
    \toprule
	\midrule
    \textbf{\textbf{\quad Constraint}} & 
    \multicolumn{3}{c}{\textbf{\textbf{Signal}}} & 
    \multicolumn{3}{c}{\textbf{\textbf{Background (Total)}}} \\ 
    \cline{2-7}
    &\quad $\sigma$ (fb)&$\varepsilon_{\text{abs}}$ &$\varepsilon_{\text{rel}}$ \quad& \quad$\sigma$ (fb)&$\varepsilon_{\text{abs}}$ &$\varepsilon_{\text{rel}}$ \quad\\ 
    \midrule
    Initial (MG output)  & 0.264 & 100\% & $\backslash$ & 1.161 & 100\% & $\backslash$ \\
    Leptons selection & 0.165 & 62.5\% & 62.5\% & 0.446 & 38.4\% & 38.4\% \\
    $M_{Z_1} \, \& \, M_{Z_2}$& 0.109 & 41.3\% & 66.1\% & 0.016 & 1.4\% & 3.6\% \\
    $M_{4\ell}$& 0.107 & 40.7\% & 98.2\% & 4.90E-6 & 0.0\% & 0.0\% \\
	\midrule
    \bottomrule
    \end{tabular}
    \caption{Conclusion of the cut flow in this study.}
    \label{tab:efficiency_table}
\end{table}

\section{Numerical results}
In this section, we present the results of the coefficient tomography and entanglement study between the two Z bosons in the signal process. In Sec.~\ref{sec:id5_1}, we use high-statistics samples to demonstrate the reconstruction of the density matrix coefficients and the expected values of $c^2_{\text{MB}}$ and $\mathcal{N}(\rho)$; In Sec.~\ref{sec:id5_2}, we analyze numerical results with statistics matching experimental expectations and evaluate the significance of QE observation. For each study, two distinct levels of sample are considered:
\begin{itemize}
    \item Truth Level (TL): Generated directly from MG at the Les Houches Event (LHE) level, without further simulation via PY8 or Delphes. On this basis, we remove all constraints on the final-state leptons in MG, thereby obtaining the samples of the full phase space.
    \item  Reconstructed Level (RL): Incorporates full simulation chains via MG, PY8, and Delphes, as well as the signal selection and background suppression. All selection criteria align with those described in Sec.~\ref{sec:sbstudy}.
\end{itemize}

\subsection{Results of the tomography and the QE observable}
\label{sec:id5_1}
\begin{figure}[h]
    \centering
    \subfloat[Truth level]
    {\includegraphics[width=.5\textwidth]{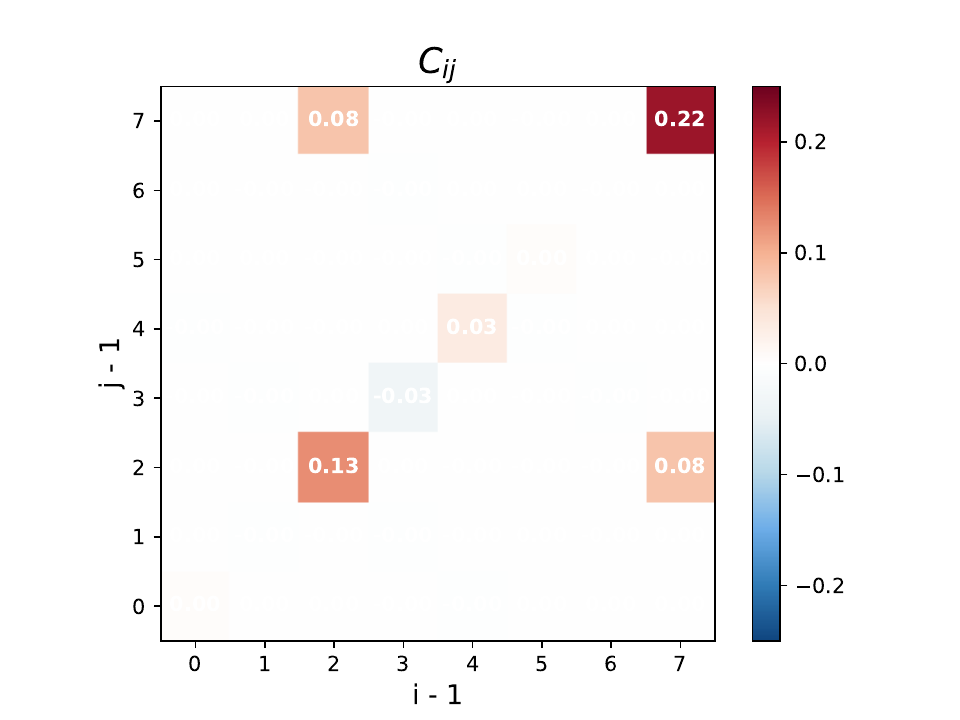}\label{fig:id3_1}}
     \subfloat[Reconstructed level]
    {\includegraphics[width=.5\textwidth]{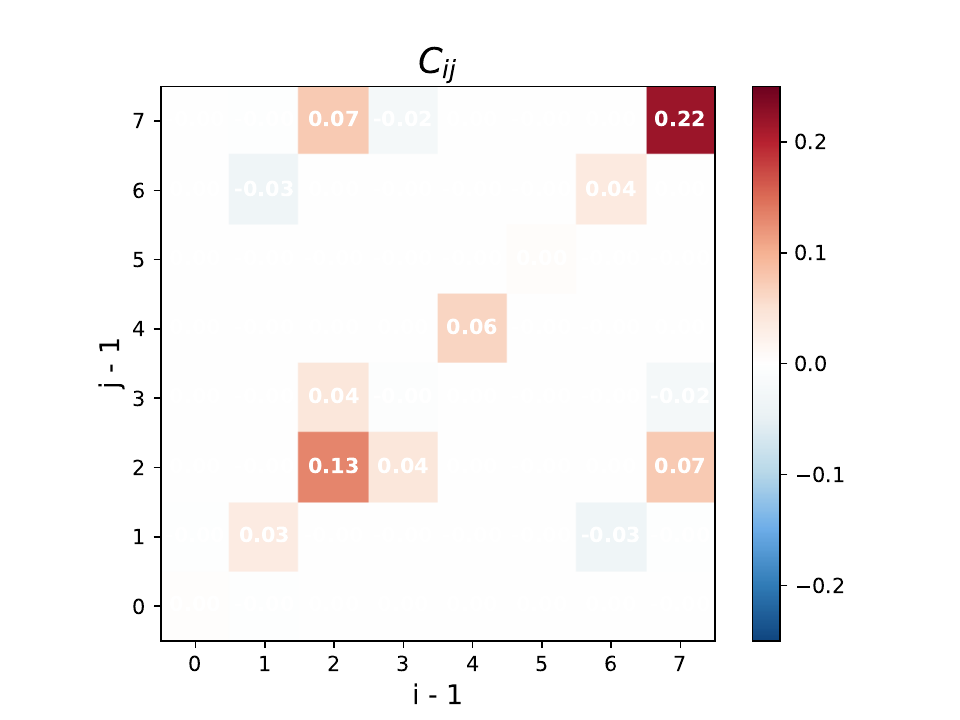}\label{fig:id3_2}}
    \caption{Coefficient $C_{ij}$ tomography results of the $ZZ$ system generated from $\mu ^+ \mu ^- \to Z Z$ process.}
    \label{fig:id3}
\end{figure}

In this section, we simulate 20 million MC events of the signal process $\mu ^+ \mu ^- \to Z Z \to 4\ell$ and apply the quantum tomography framework described in Sec.~\ref{sec:tomography} to reconstruct the quantum state with high statistics. The TL analysis utilizes the full simulated sample, while the RL retains approximately 8 million events after incorporating detector response simulations and kinematic selections, consistent with the signal efficiency in Sec.~\ref{sec:sbstudy}. The extracted coefficient $C_{ij}$ of the density matrix defined in Eq.~\ref{eq:generaldec} are shown in Fig.~\ref{fig:id3}. 

According to the results, the density matrix decomposition at TL exhibits significant non-zero coefficients $C_{33}$, $C_{38}$, $C_{83}$, $C_{88}$, $C_{44}$, and $C_{55}$, with other components being either strictly zero or strongly suppressed due to dependencies on the center-of-mass energy to Z boson mass ratio~\cite{Fabbrichesi:2023cev}. Meanwhile, the dominant coefficients ($C_{33}$, $C_{38}$, $C_{83}$ and $C_{88}$) exhibit agreement between TL and RL samples, but non-negligible deviations emerge in other components, primarily induced by the phase-space constraints from detector simulations and kinematic selections. In experimental analyses, detector-level reconstructions typically require unfolding techniques to mitigate resolution effects and recover TL observables. However, this study focuses on fundamental quantum correlations, thus deferring systematic treatments to future work.

Based on the reconstructed coefficients, we can calculate the QE observables $c^2_{\text{MB}}$ and $\mathcal{N}(\rho)$ follow Sec.~\ref{sec:observables}, as
\begin{equation}
    \label{eq:c2_results_mum}
    \begin{aligned}
    c^2_{\text{MB,TL}} &= 0.077,\quad c^2_{\text{MB,RL}} = 0.179,\\
    \mathcal{N}_{\text{TL}}&=0.093,\quad \mathcal{N}_{\text{RL}}=0.208.
    \end{aligned}
\end{equation}
As described in Sec.~\ref{sec:observables}, values $c^2_{\text{MB}}>0$ or $\mathcal{N} >0$ indicate the presence of entanglement in the system.

\subsection{Entanglement observation at experimental scenario}
\label{sec:id5_2}
In actual collider experiments, the event counts are orders of magnitude lower than those in the high-statistics simulations of Sec.~\ref{sec:id5_1}. To systematically investigate the prospective outcomes and the statistical significance, we evaluate a integrated luminosity range of 5~ab$^{-1}$ to 75~ab$^{-1}$. The expected event yields depend on both the signal efficiency and production cross-section derived in Sec.~\ref{sec:sbstudy}. For the TL samples, while no constraints are applied during generation, we apply the RL signal efficiency to these samples to ensure consistency with realistic experimental conditions.

\begin{figure}[htbp]
	\centering
	\subfloat[$C_{33}$]
	{\includegraphics[width=.45\textwidth]{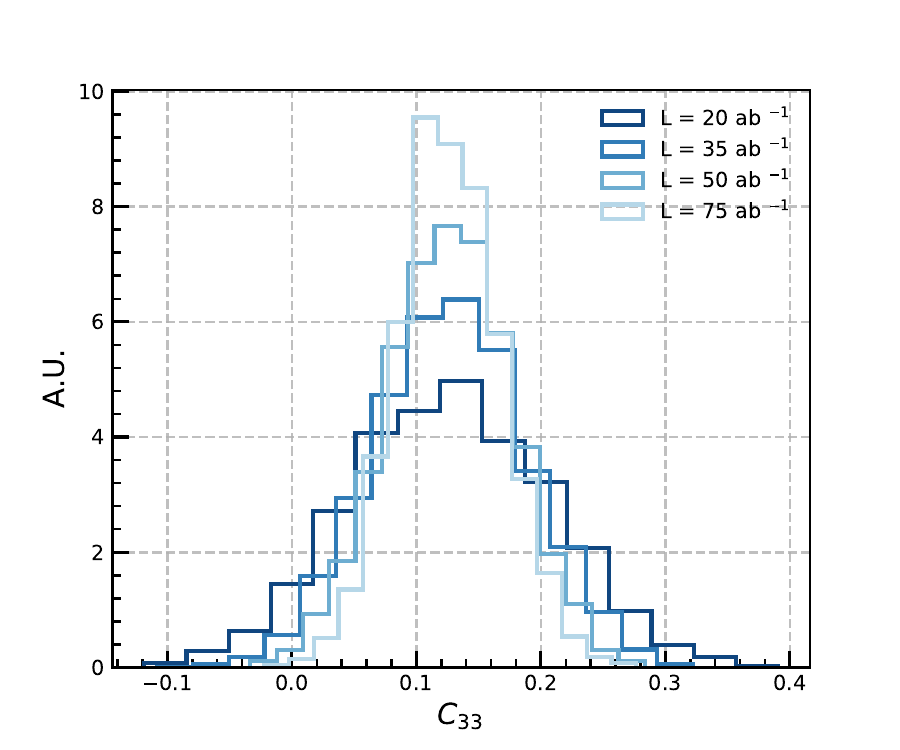}}
	\subfloat[$C_{38}$/$C_{83}$]
	{\includegraphics[width=.45\textwidth]{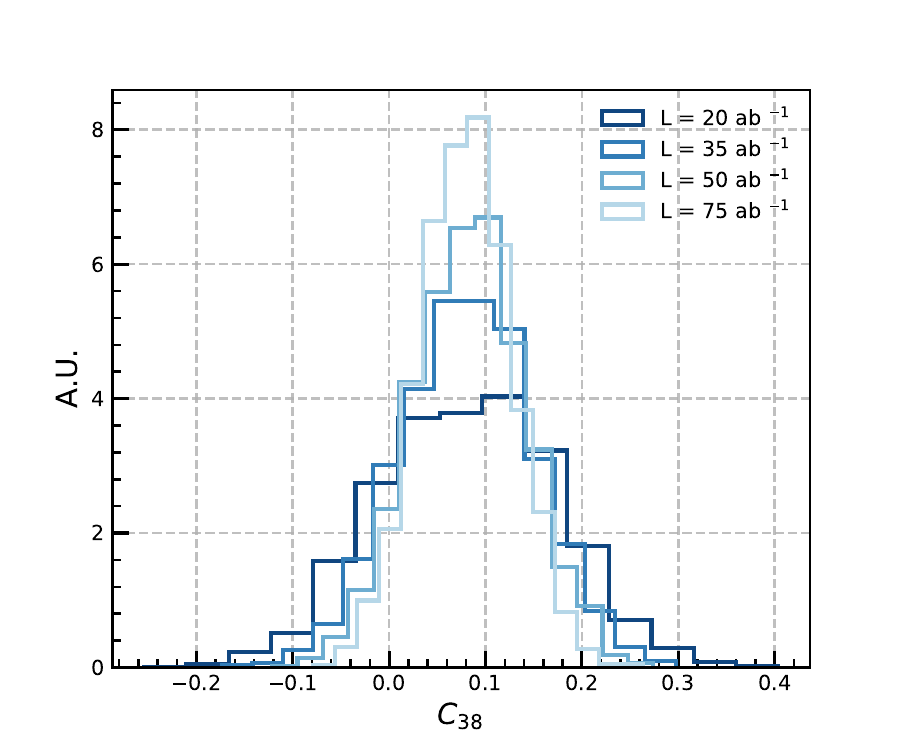}}\\
	\subfloat[$C_{88}$]
	{\includegraphics[width=.45\textwidth]{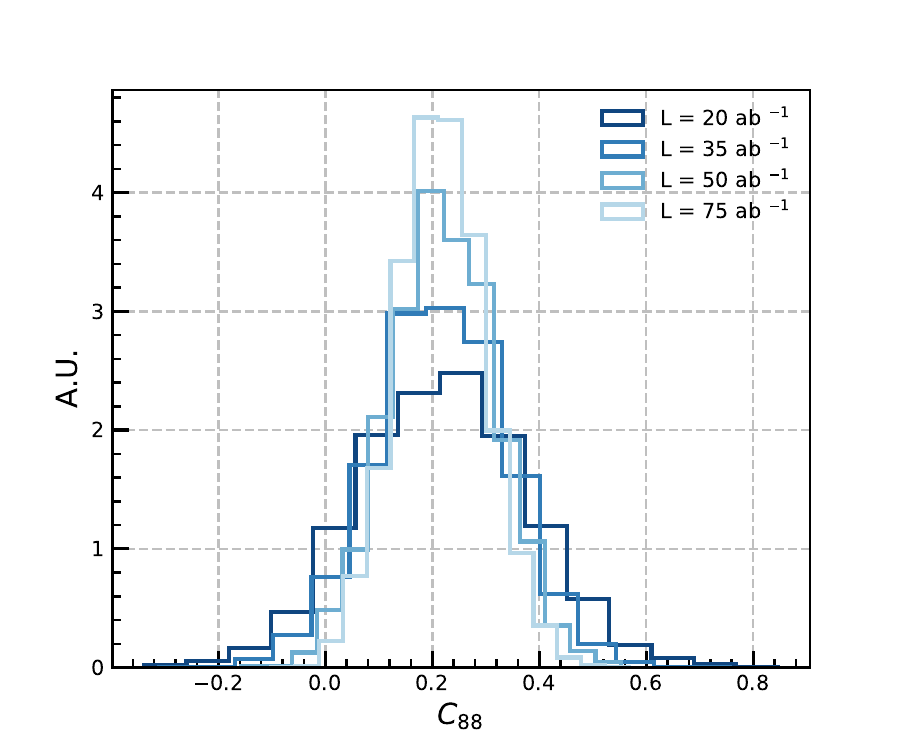}}
	\subfloat[$C_{44}$]
	{\includegraphics[width=.45\textwidth]{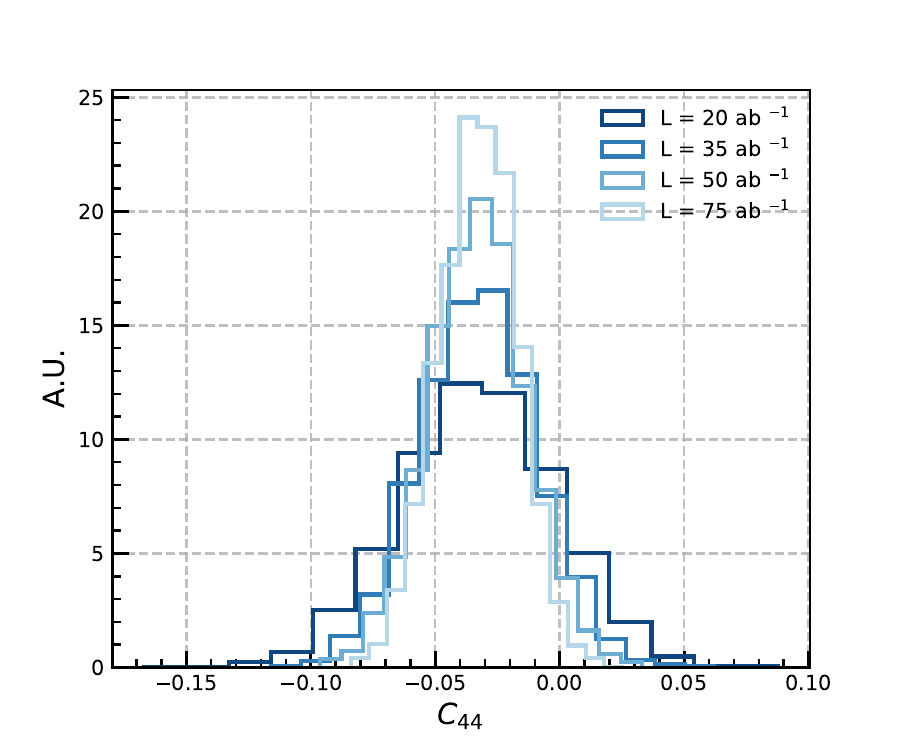}}
	\caption{Distributions of coefficients $C_{33}$, $C_{38}$, $C_{88}$, and $C_{44}$ from 4,000 pseudo experiments under different luminosity at TL.}
	\label{fig:id4}
\end{figure}

Using the high-statistics sample generated in Sec.~\ref{sec:id5_1}, we extract 4,000 small samples of each event yields, assuming each sample as one pseudo experiment. For each pseudo experiment, we compute the density matrix coefficients. The resulting distributions are shown in Fig.~\ref{fig:id4}.
All coefficient distributions exhibit approximately Gaussian profiles, with central values are consistent with the high-statistics results in Fig.~\ref{fig:id3}, as expected.

\begin{figure}[htbp]
	\centering
	\subfloat[$c^2_{\text{MB}}$]
	{\includegraphics[width=.49\textwidth]{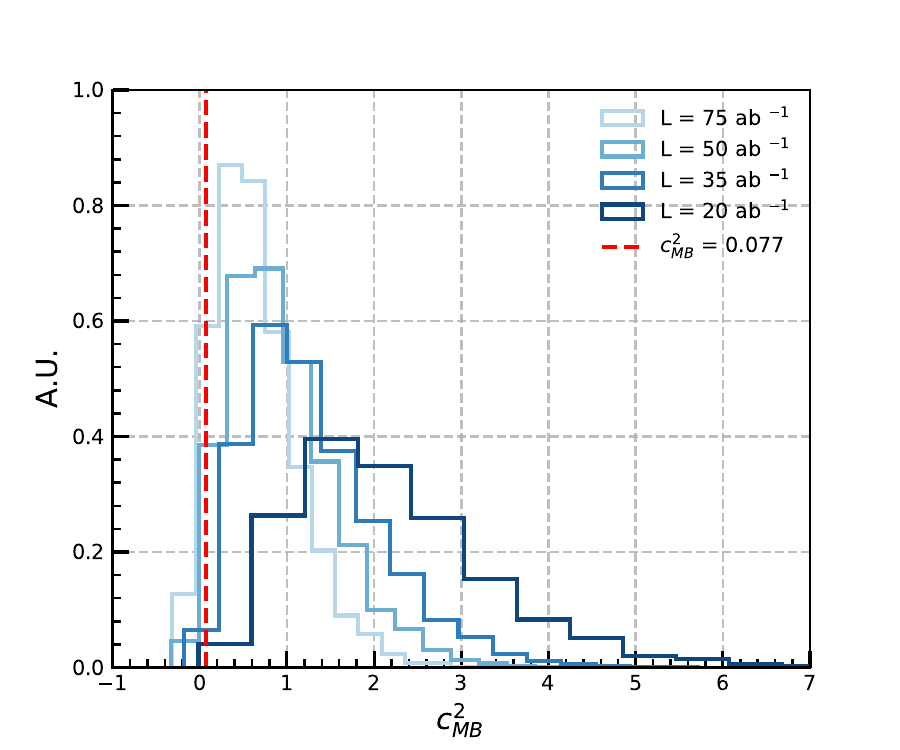}}
	\subfloat[$\mathcal{N}$]
	{\includegraphics[width=.49\textwidth]{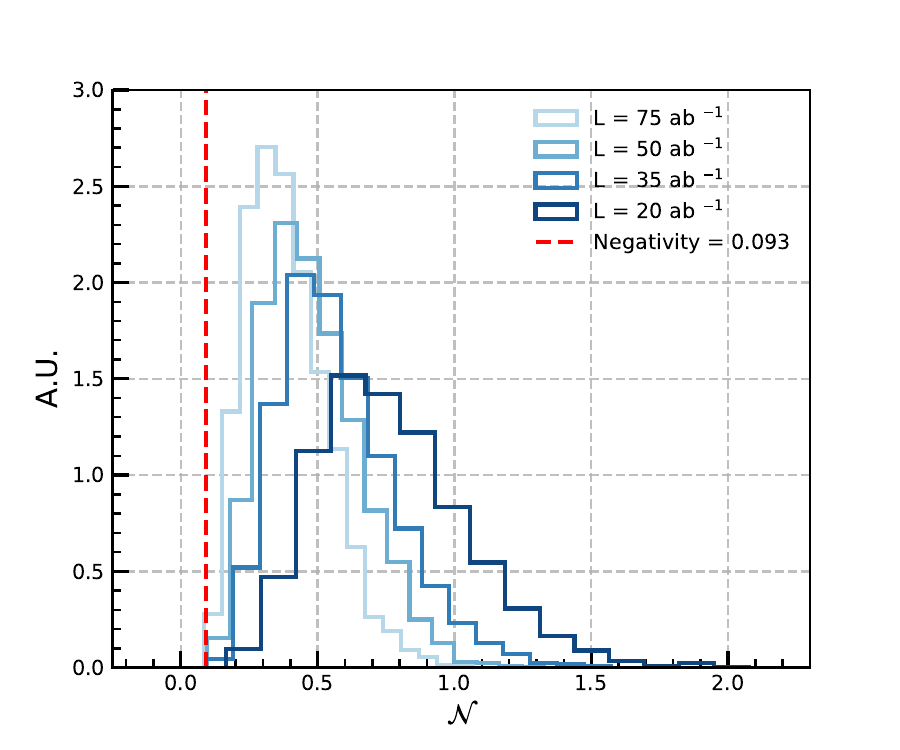}}
	\caption{Distributions of QE observables $c^2_{\text{MB}}$ and $\mathcal{N}$ from 4,000 pseudo experiments under different luminosity at TL. The red dashed lines demonstrate the expected value given by large samples as Eq.~\ref{eq:c2_results_mum}.}
	\label{fig:dis_c2_n}
\end{figure}

Next, we focus on the distributions of the QE observables. As shown in Fig.~\ref{fig:dis_c2_n}, both $c^2_{\text{MB}}$ and $\mathcal{N}$ exhibit significantly skewed distributions across the studied luminosity range. Notably, the expectation values display a monotonic increase with decreasing luminosity, deviating substantially from the high-statistics baseline values in Eq.~\ref{eq:c2_results_mum}.

\begin{figure}[htbp]
	\centering
	\subfloat[Truth level]
	{\includegraphics[width=.5\textwidth]{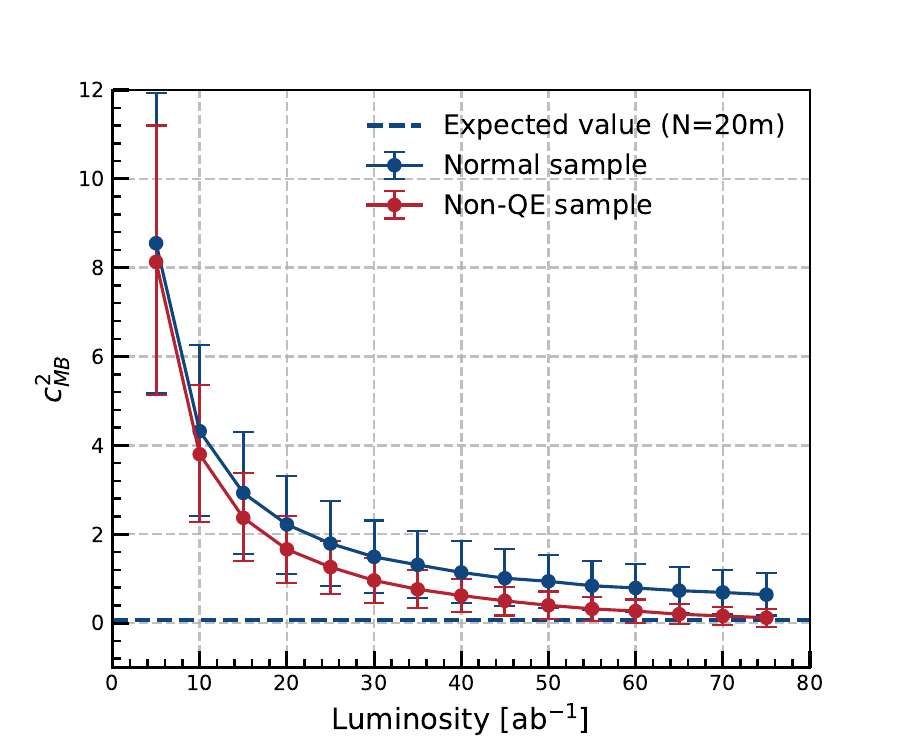}}
	\subfloat[Reconstructed level]
	{\includegraphics[width=.5\textwidth]{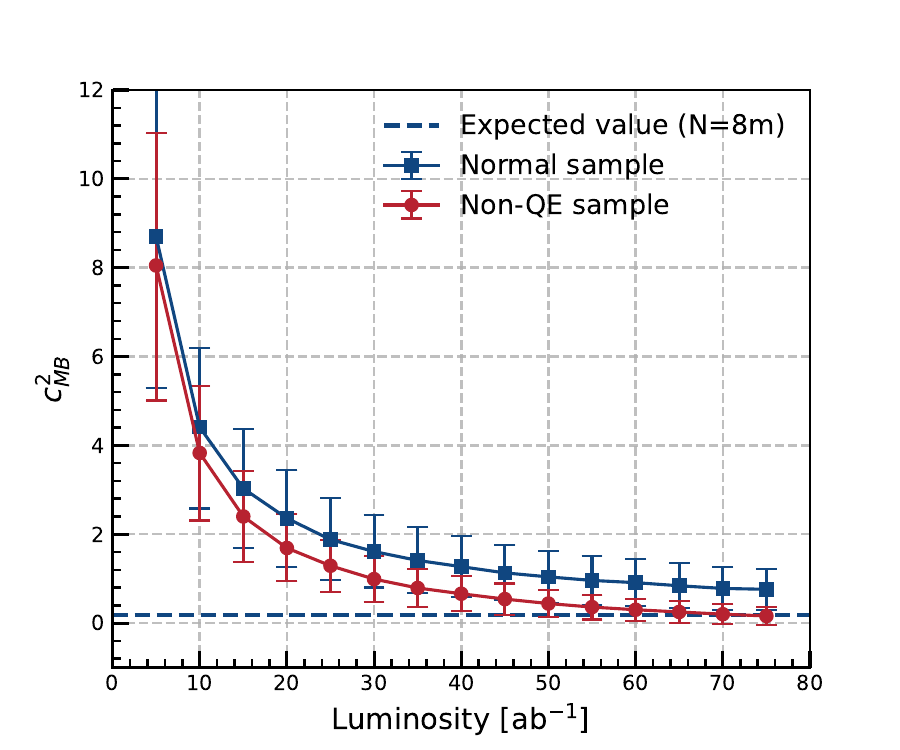}}
	\caption{Comparison of $c_{\text{MB}}^2$ value between the normal samples and the non-QE samples of different luminosity, where the points are the expectations and the error bars are both within the 1$\sigma$ (68.3\%) equal-tailed intervals, and the dashed lines represent the expected values obtained by the large samples as Eq.~\ref{eq:c2_results_mum}.}
	\label{fig:id5}
\end{figure}

Given the observed luminosity-dependent biases in QE observables, characterizing entanglement significance via expectation values and standard deviations becomes inadequate. Therefore, we implement a hypothesis testing framework using non-entangled (non-QE) samples.
The non-QE samples are simulated by using Madspin~\cite{Artoisenet:2012st} within the MG workflow, which allows us to get the $ZZ$ events without spin correlation. This preserves classical correlations (e.g., momentum conservation) while eliminating quantum entanglement.
The non-QE density matrix reduces to a normalized 9-dimensional identity matrix, with $c^2_{\text{MB}}= -4/9$ and $\mathcal{N}=0$. Then we obtain 4,000 non-QE samples with varying sample sizes, following the same procedure as the normal sample analysis described above. Similar to the behavior observed in the normal samples, non-QE samples exhibit rising $c^2_{\text{MB}}$ and $\mathcal{N}$ expectations as luminosity decreases. In Fig.~\ref{fig:id5} it shows the comparison of $c_{\text{MB}}^2$ value between the normal samples and the non-QE samples, revealing that the statistical divergence between the normal samples and the non-QE samples grows with increasing luminosity.

\begin{figure}[!h]
    \centering
    \subfloat[p-value calculation method]
	{\includegraphics[width=.5\textwidth]{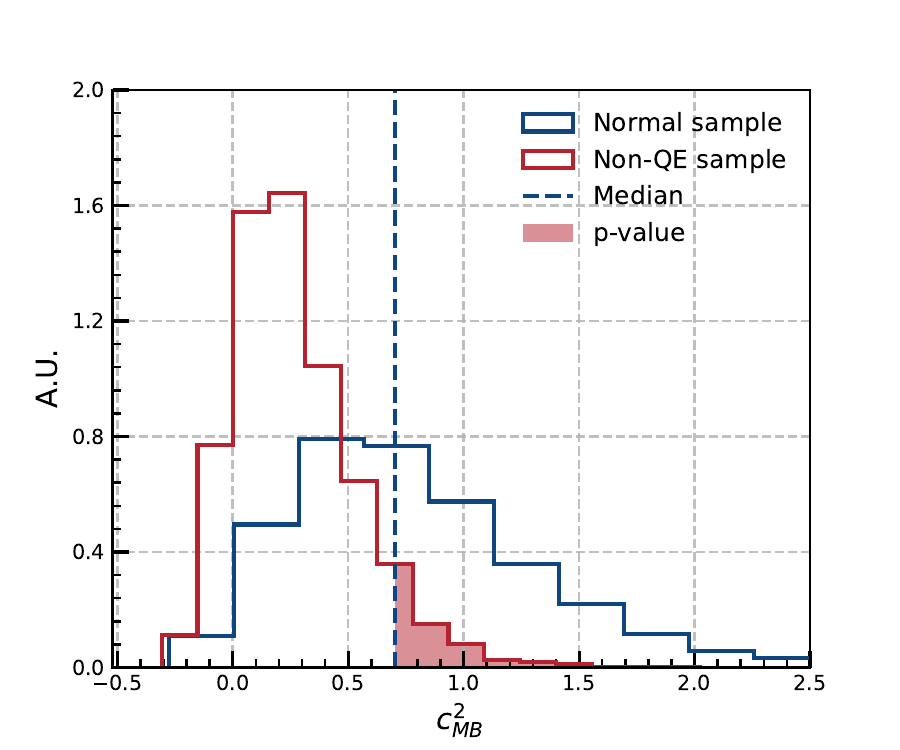}\label{fig:id6_1}}
    \subfloat[Significance results]
	{\includegraphics[width=.5\textwidth]{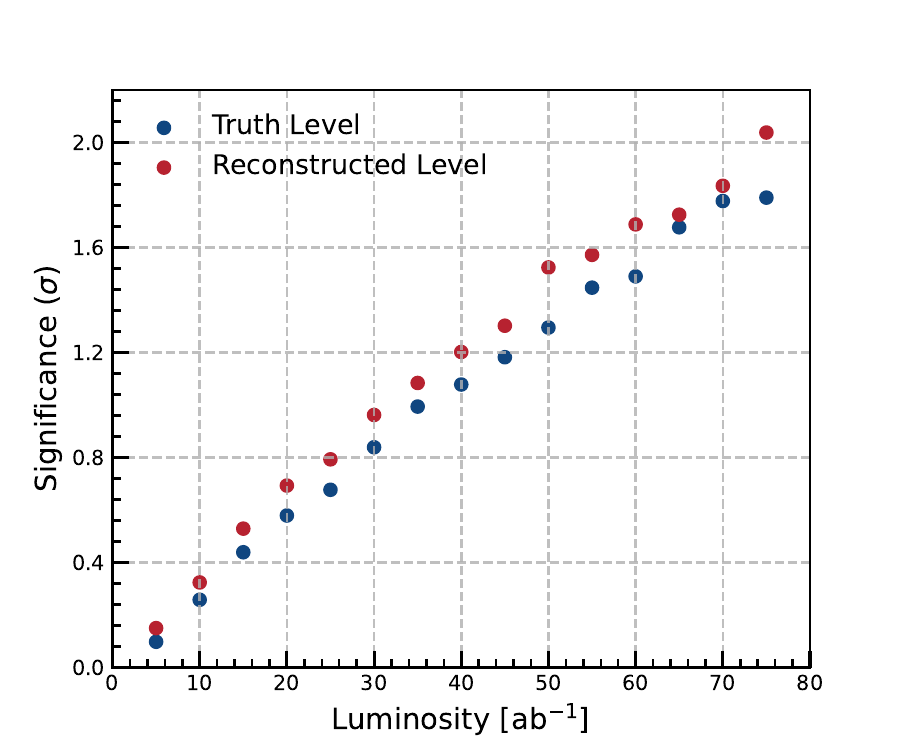}\label{fig:id6_2}}
    \caption{Schematic diagram of the p-value calculation method (left), and the significance results (right).}
    \label{fig:id6}
\end{figure}

To quantify the statistical significance of QE, we perform hypothesis testing against the non-QE null hypothesis. Considering the non-QE sample distribution as the null hypothesis, we calculate the p-value for each pseudo experiment and obtain the p-value distribution. Given the asymmetry of the p-value distribution, we ultimately adopt the median value as the representative result for a real experiment. This approach is mathematically equivalent to a single hypothesis test using the median $c_{\text{MB}}^2$ value across pseudo-experiments. The schematic diagram are illustrated in Fig.~\ref{fig:id6_1}, taking $L=60~$ab$^{-1}$ at TL as an example. Figure~\ref{fig:id6_2} converts the median p-values into Gaussian significances (Z-scores, or $\sigma$) via the relation $Z = \Phi^{-1}(1-p)$, where $\Phi^{-1}$ is the inverse standard normal cumulative distribution function.

The results indicate that QE can be observed with a statistical significance of 2$\sigma$ when the integrated luminosity reaches approximately 80 ab$^{-1}$. However, under low-statistics conditions, the significance deteriorates rapidly due to the limited entanglement degree of the system and the dominance of statistical fluctuations. Theoretically, enhancing the system’s entanglement degree through phase-space constraints (e.g., restricting scattering angles $\Theta$) is possible. As demonstrated in Ref.~\cite{Fabbrichesi:2023cev}, we can restrict the scattering polar angle to $\cos \Theta$ approaches 0 to achieve higher entanglement degree. However, such constraints conflict with the angular distribution of the scattering process - applying such constraints would drastically reduce signal efficiency, rendering this approach nonviable upon comprehensive evaluation of the statistic. 

Notably, aside from the Higgs-mediated $ZZ$ system, most common di-boson systems produced in the collider experiment exhibit inherently low entanglement degree due to their production mechanisms~\cite{Fabbrichesi:2023cev,Morales:2023gow}. 
This fundamental limitation underscores the necessity of advancing both experimental capabilities and analytical methodologies to probe quantum correlations in high-energy collisions. We look forward to the enhancement technological upgrades in experimental capabilities and luminosity, which will directly enhance measurement precision. Additionally, it deserves a further study on exploring innovative observation frameworks to develop characterization methodologies that exhibit enhanced sensitivity and significance to quantum mechanical properties of the particle systems at high-energy collider.

\section{Conclusion}
This investigation focuses on reconstructing the spin density matrix and analyzing quantum entanglement in the $ZZ$ boson system via the process $\mu ^+ \mu ^- \to Z_1 Z_2,~ Z_1 \to \ell^+_1\ell^-_1,~Z_2 \to \ell^+_2\ell^-_2$ with $\ell=e,\mu$ at a 1~TeV muon collider. Through systematic MC simulations and background suppression procedure, we demonstrate that residual background contributions are negligible, allowing their exclusion from the final analysis. Utilizing a high-statistics sample of 20 million simulated signal events, we achieve full reconstruction of the density matrix coefficients via quantum state tomography, with the measurement of QE observable $c^2_{\text{MB}}$ and $\mathcal{N}$ revealing entanglement signatures in the $ZZ$ system. By evaluating expected signal event yield across an integrated luminosity range of 5~ab$^{-1}$ to 75~ab$^{-1}$, we obtain 4,000 pseudo-experiments, giving the histograms of the coefficients and observables of different luminosity.

The study conclusively shows that under low-statistics conditions, QE observables exhibit pronounced asymmetric distributions. Consequently, conventional evaluation framework, which reliants on "mean $\pm$ standard deviation" approximations and inadequate pseudo-experiment sampling, will systematically overestimate the significance of QE observation in such $\mu\mu\to ZZ$ processes. To address this bias, we generate spin-uncorrelated events, namely the non-QE samples.
By performing hypothesis testing against this null hypothesis, we compute median p-values across 4,000 pseudo-experiments and convert them into Gaussian significances, providing a statistical reference for experimental practice. In conclusion, QE can be observed with a statistical significance of 2$\sigma$ when the integrated luminosity reaches approximately 80 ab$^{-1}$.

{
\appendix
\section{The analytical expressions of $p_i^+$ and matrix $A$}
According to Ref.~\cite{Ashby-Pickering:2022umy}, the analytical expressions of $p_i^+$ in Eq.~\ref{eq:calc_a_i} read
\begin{eqnarray}
    \label{eq:def_p_plus}
    \begin{aligned}
        p_1^\pm &= \sqrt{2} \sin \theta \, (5 \cos \theta \pm 1) \cos \phi ,\\
        p_2^\pm &= \sqrt{2} \sin \theta \, (5 \cos \theta \pm 1) \sin \phi ,\\
        p_3^\pm &= \left(5 \pm 4 \cos \theta + 15 \cos 2\theta\right)/ 4 ,\\
        p_4^\pm &= 5 (\sin \theta)^2 \cos 2\phi ,\\
        p_5^\pm &= 5 (\sin \theta)^2 \sin 2\phi ,\\
        p_6^\pm &= \sqrt{2} \sin \theta \, (-5 \cos \theta \pm 1) \cos \phi ,\\
        p_7^\pm &= \sqrt{2} \sin \theta \, (-5 \cos \theta \pm 1) \sin \phi ,\\
        p_8^\pm &= \left(-5 + 12 \cos \theta - 15 \cos 2\theta\right) / 4 \sqrt{3},
        \end{aligned}
\end{eqnarray}
where $\phi$ and $\theta$ are the azimuth and polar angle of the decay lepton $\ell^+$ in its mother's rest frame, respectively.

And the matrix $A$ in Eq.~\ref{eq:calc_p_tilde} reads
\begin{eqnarray}
    \label{eq:def_A_matrix}
    A = \frac{1}{g^2_R-g^2_L}
\begin{bmatrix}
g_R^2 & 0 & 0 & 0 & 0 & g_\ell^2 & 0 & 0 \\
0 & g_R^2 & 0 & 0 & 0 & 0 & g_\ell^2 & 0 \\
0 & 0 & g_R^2 - \frac{1}{2} g_\ell^2 & 0 & 0 & 0 & 0 & \frac{\sqrt{3}}{2} g_\ell^2 \\
0 & 0 & 0 & g_R^2 - g_\ell^2 & 0 & 0 & 0 & 0 \\
0 & 0 & 0 & 0 & g_R^2 - g_\ell^2 & 0 & 0 & 0 \\
g_\ell^2 & 0 & 0 & 0 & 0 & g_R^2 & 0 & 0 \\
0 & g_\ell^2 & 0 & 0 & 0 & 0 & g_R^2 & 0 \\
0 & 0 & \frac{\sqrt{3}}{2} g_\ell^2 & 0 & 0 & 0 & 0 & \frac{1}{2} g_\ell^2 + g_R^2
\end{bmatrix},
\end{eqnarray}
where $g_R$ and $g_L$ are the right- and left-chiral couplings.
}

\bibliographystyle{JHEP}
\bibliography{biblio.bib}






\end{document}